# Controllable creation of topological boundary states in topological-insulator-based Josephson corner junctions

Xiang Wang[1,2], Duolin Wang[1,2], Yunxiao Zhang[1,2], Xiaozhou Yang[1,2], Yukun Shi[1,2], Bing Li[1,2], Enna Zhuo[1,2], Yuyang Huang[1,2], Anqi Wang[1,2], Zhaozheng Lyu[1,2,3]†, Xiaohui Song[1,3], Peiling Li[1], Bingbing Tong[1], Ziwei Dou[1], Jie Shen[1], Guangtong Liu[1,3], Fanming Qu[1,3,4] and Li Lu[1,3,4]†

[1] *Beijing National Laboratory for Condensed Matter Physics, Institute of Physics, Chinese Academy of Sciences, Beijing 100190, China*

[2] *School of Physical Sciences, University of Chinese Academy of Sciences, Beijing 100049, China*

[3] *Hefei National Laboratory, Hefei 230088, China*

[4] *Songshan Lake Materials Laboratory, Dongguan, Guangdong 523808, China*

† Corresponding authors: lyuzhzh@iphy.ac.cn, lilu@iphy.ac.cn

Abstract

Majorana zero modes (MZMs) in condensed matter systems have attracted great attention in the past two decades, due to their interesting physics and potential application in topological quantum computing (TQC). However, the topologically protected nature of MZMs still need more experimental verifications. In this study, we have realized controllable creation of a topological boundary state at the corner of topological insulator (TI)-based Josephson corner junctions. This state demonstrates protected existence across a broad region in parametric space, and exhibits a non-$2\pi$-period but $4\pi$-period-compatible energy-phase relation. Our study suggests that TI-based Josephson junctions, as proposed in the Fu-Kane scheme of TQC, may provide a promising platform for hosting and braiding MZMs.

MZMs as a type of topological boundary state were predicted to exist at the ends of 1D spinless *p*-wave superconductors [1], in devices composed of *s*-wave superconductors and semiconductor nanowires with strong spin-orbit coupling [2,3], in Josephson junctions (JJs) or vortex cores of *p*-wave-like superconductors formed of *s*-wave superconductors and TIs [4,5], etc. With these MZMs, various schemes for realizing TQC have been proposed, such as the one based on nanowires [6,7], and the one based on TIs [5,8]. In the latter, known as the Fu-Kane scheme, TI-based Josephson trijunctions are proposed to serve as the fundamental building blocks of topological quantum circuits [5], and surface codes are utilized to enable universal TQC [8].

Experimentally, signatures of MZMs such as zero-bias conductance peaks (ZBCPs) have been observed in various systems, including nanowire devices [9-11], atomic [12,13] and artificial [14] chains, superconductor-TI interfaces [15,16], vortex cores of TI-based [17,18] and Fe-based superconductors [13,19-21]. Additionally, a $4\pi$-period current-phase relation (CPR) or energy-phase relation (EPR) in JJs made of *p*-wave or *p*-wave-like superconductors [1,5,22] has also been observed, signified as fractional Josephson effect in nanowire-based [23] and TI-based Josephson junctions (TIJJs) [24], and as the linear and complete closure of minigap in TIJJs [25,26], where the minigap is defined as half of the energy spacing between the lowest electron-like and hole-like Andreev bound states. The minigap closure has also been studied in 2D electron gas-based JJs in conjunction with ZBCP measurement [27,28]. Despite these advances, however, the search for robust MZMs that can be controllably created and conveniently braided continues.

As topological boundary states protected by global invariants, MZMs are expected to exhibit robustness against local perturbations, persisting across broad parametric ranges similar to quantum Hall edge states [29]. To probe the controllable creation and stability of such states in TIJJs, we fabricated Al-$Bi_2Te_3$-Al Josephson corner junctions on $Bi_2Te_3$ surfaces [Fig. 1(a)]. These perpendicular x/y-junctions enable independent control of anomalous phase shifts by using an in-plane magnetic field via anomalous Josephson effect [30-35]. At T≈30 mK, we demonstrated a robust topological corner state with full local minigap closure, showing $4\pi$-period-compatible behavior and broad parametric stability.

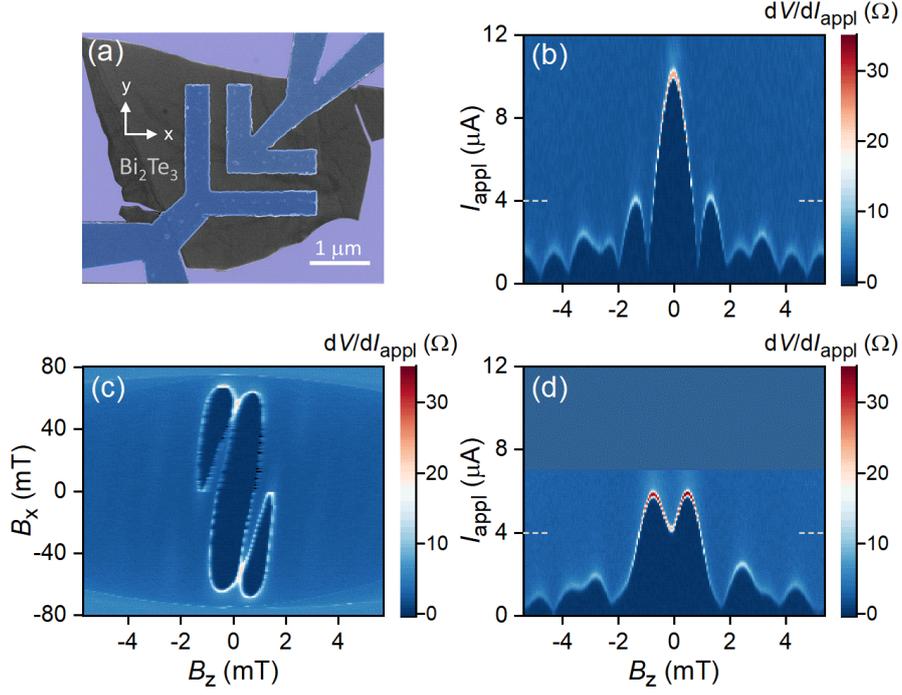

FIG. 1. (a) Scanning electron micrograph (SEM) of Al-Bi$_2$Te$_3$-Al corner junction (Bi$_2$Te$_3$: 20-30 nm thick; superconducting electrodes: 5 nm Ti/50 nm Al). (b) Zero-resistance lobes vs. applied current $I_{appl}$ and out-of-plane magnetic field $B_z$ (at in-plane magnetic field $B_x$=0). (c) $B_x$-induced shifts of the central three lobes measured at $I_{appl}$=4 μA. (d) Zero-resistance lobes measured at $B_x$=52 mT, reflecting that the phase difference jump at the corner reached π.

Firstly, we measured the critical supercurrent of the corner junction in an out-of-plane magnetic field $B_z$ and in the absence of an in-plane magnetic field. A standard Fraunhofer pattern was observed [Fig. 1(b)]. Its oscillation period of approximately 0.9 mT matches with the expected one of $\Delta B_z = \phi_0/2WL^* = 0.84$ mT, where $\phi_0$ is the flux quantum, $W = 1.76$ μm is the width of the line junctions, and $L^* = 0.7$ μm is the effective length of the junction between the centers of the two superconducting electrodes.

Secondly, by inducing an anomalous phase shift $\varphi_{Ax}$ in x-junction through the application of $B_x$, a phase difference jump $\Delta\varphi = \varphi_{Ax}$ occurred at the corner, accompanied with an internal circulating supercurrent that flows between the two line junctions. This phase difference jump reduced the total critical supercurrent, shifting the lobes with $B_x$ as depicted in Fig. 1(c). When $\Delta\varphi = \varphi_{Ax} = \pi$, the total critical supercurrent of the corner junction was minimized at $B_z=0$ [Fig. 1(d)]. The residual supercurrent at $B_z=0$ arose because the two line junctions were asymmetrically affected when the in-plane magnetic field was applied along only one of them. To ensure that the two line junctions are symmetrically influenced other than in their anomalous phases, we chose to apply the in-plane magnetic field along ±45° directions for the remainder of the experiment. By doing so, the total critical supercurrent at $B_z=0$ could indeed be suppressed to zero [36].

In the next, we integrated a superconducting loop onto the corner junction [Figs. 2(a), 2(b)] to generate a uniform phase difference $\varphi$ via an out-of-plane magnetic flux. Additionally, three normal-metal probe electrodes were fabricated to monitor the local minigap's opening and closing in the line junctions and at the corner through contact resistance measurement.

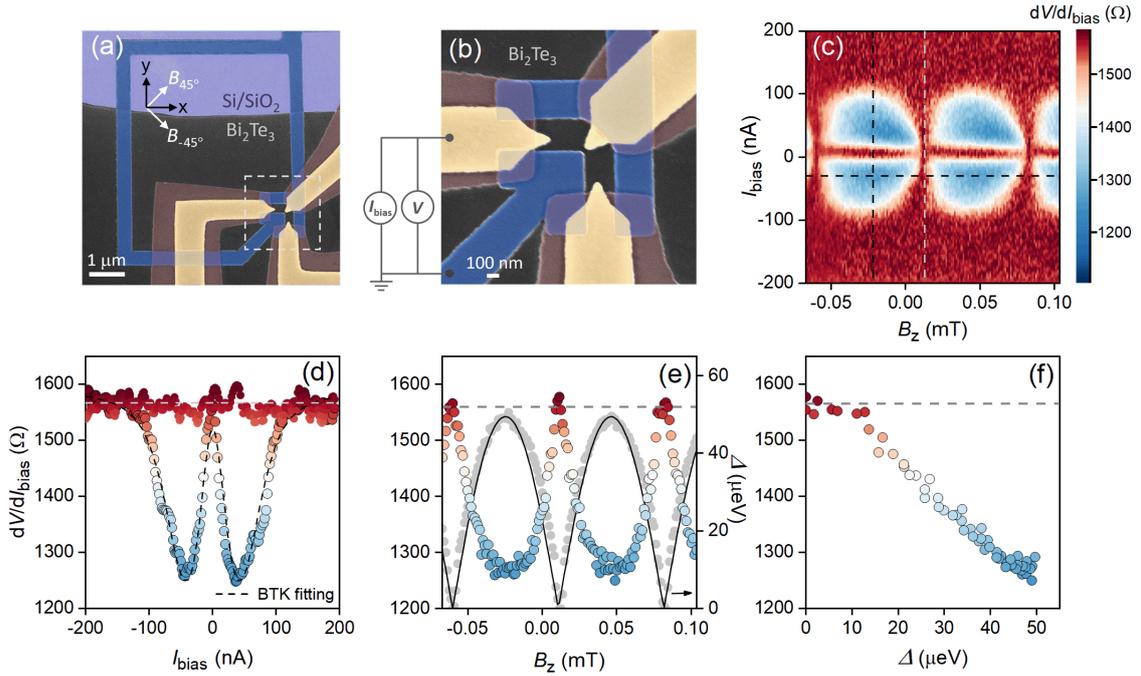

FIG. 2. (a, b) False-color SEM images of the device. A superconducting loop (blue) was connected to a corner junction. Three normal-metal probes (yellow, 2 nm Al/100 nm Au) were fabricated to detect the local minigap at the corner and the centers of line junctions via contact resistance measurement. $Al_2O_3$ pads (brown, 30 nm) were used to isolate the probes, allowing contact only at their tips. (c) 2D contact resistance map measured at the center of x-junction under a 42.4 mT in-plane field at -45°. (d) Vertical linecuts in (c) at $B_z=-0.0216$ mT and 0.0126 mT, with BTK fitting to one of them (black dashed line). Gray dashed line indicates normal-

state resistance. (e) Horizontal linecut in (c) at $I_{bias}$=-30 nA, with minigap (gray dots) extracted from (c). The Black line shows $|\cos(\varphi/2)|$, with a $2\pi$ periodicity of 0.072 mT determined by the area of the superconducting loop (~30 μm²). (f) Correspondence between contact resistance and minigap in (e).

Figure 2(c) is the 2D map of the contact resistance measured at the center of x-junction as a function of $I_{bias}$ and $B_z$. It visually depicts the opening and closing of the local minigap when the magnetic flux through the superconducting loop is varied.

In Fig. 2(d), we present the contact resistance data along the two vertical linecuts indicated in Fig. 2(c), together with the Blonder-Tinkham-Klapwijk (BTK) fitting [37,38] through which the minigap at corresponding magnetic fields can be extracted [36]. The maximum minigap, $\Delta_{max}$, is found to be 48.5 μeV at $B_z$=-0.0216 mT. Conversely, the data along the other linecut at $B_z$=0.0126 mT exhibits a gapless feature, indicating that the minimum minigap, $\Delta_{min}$, is essentially zero.

In Fig. 2(e), we show the contact resistance along the horizontal linecut in Fig. 2(c) and the minigap extracted from vertical linecuts at each $B_z$. The minigap follows $|\cos(\varphi/2)|$, with $\Delta_{min} < 1$ μeV at $\pi$ phase difference. In superconductor-normal metal-superconductor (S-N-S) JJs, this form can arise in the diffusive transport limit with high density of states [39] or in the short-junction limit with near-perfect transmission [40]. For our S-TI-S JJs, the transmission coefficient $\tau = 1-(\Delta_{min}/\Delta_{max})^2 = 1-(1\text{ μeV}/48.5\text{ μeV})^2 > 0.999$, approaching $\tau \to 1$, leading to $\Delta \propto \sqrt{1-\tau\sin^2(\varphi/2)} | = |\cos(\varphi/2)|$ [40]. This $2\pi$-period form could also stem from a topologically nontrivial $4\pi$-period $\cos(\varphi/2)$ [1,5,22], due to the single-helicity nature of electrons on TI surface where electron backscattering via ground Andreev bound states is completely prohibited. While Fig. 2(e) alone cannot distinguish between $2\pi$-period and $4\pi$-period forms of minigap, we later show that the regional closure of minigap in Fig. 3(c) requires the topologically nontrivial scenario.

Figure 2(e) shows that the minigap's opening/closing is highly correlated to the local contact resistance, which is further illustrated in Fig. 2(f). This allows monitoring the minigap by tracking the local contact resistance alone. To avoid the zero-bias resistance anomaly which, according to the BTK formalism, appears at a medium-high interfacial barrier $Z$ ($Z$=0.7-1.2 in this study [36]), contact resistance was measured at a finite bias current rather than at $I_{bias}$=0.

We applied an in-plane magnetic field to induce anomalous phase shifts in the two line junctions. When applied at 45° [Figs. 3(a)-(c)], the phase shifts in the two line junctions were opposite ($\varphi_{Ax} = \varphi_A$, $\varphi_{Ay} = -\varphi_A$), creating a phase difference jump of $\Delta\varphi = 2\varphi_A$ at the corner. Conversely, at -45° [Figs. 3(d)-(f)], the phase shifts in the two line junctions were identical, resulting in no phase difference jump ($\Delta\varphi = 0$) at the corner.

Figure 3 shows the 2D maps of contact resistance versus in-plane ($B_{\pm 45°}$) and out-of-plane ($B_z$) magnetic fields. The green (yellow) dots mark the peak positions for the x(y)-junction, where the minigap is presumably closed due to $B_z$-induced sign changes [25,26]. Irregular pattern bending in the maps stems from incomplete compensation of unintended out-of-plane field component caused by sample-magnet misalignment [36].

As $B_{45°}$ increases, thus the anomalous phase shifts increase, the minigap closure positions of the two line junctions shift oppositely along the $B_z$ direction in Figs. 3(a) and 3(b). Figure 3(c) reveals that these shifts create high-resistance plateaus (dark red) in the corner's contact resistance map, indicating a regional minigap closure behavior. With further increasing $B_{45°}$, these dark red regions expand and converge at ~100 mT. At this field strength, the anomalous phase shifts $\varphi_{Ax}$ and $\varphi_{Ay}$ in the x- and y-junctions must reach ±π, respectively, and the phase difference jump $\Delta\varphi$ at the corner must be 2π, in order to give rise to the observed 2π periodicity along $B_z$ in Fig. 3(c). Notably, the 2π periodicity along the abscissa in Fig. 3(c) is not maintained - the contact resistance at $B_{45°} \approx 100$ mT does not repeat its value at $B_{45°} = 0$.

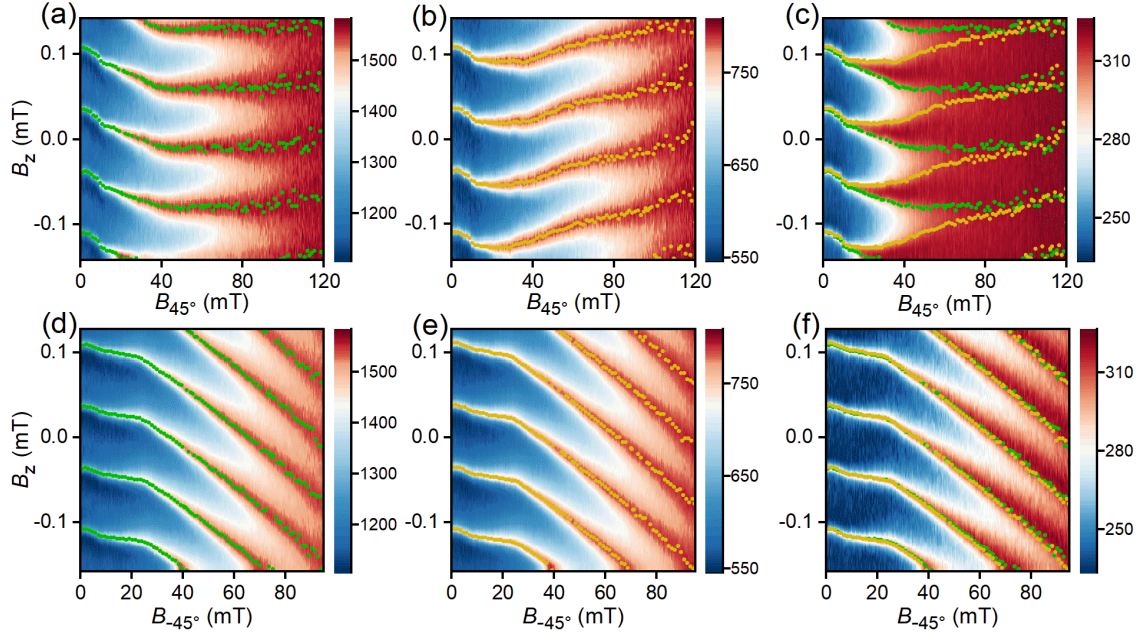

FIG. 3. Local contact resistance measured by using the normal-metal probe electrodes, as a function of out-of-plane magnetic field $B_z$ and in-plane magnetic field at the 45° (a, b, c) and -45° (d, e, f) directions. Measurements were taken at the center of the x-junction (a, d), y-junction (b, e), and corner (c, f), with $I_{bias}=-30$ nA, $-75$ nA, and $-150$ nA, respectively. Color scales indicate resistance in Ohms. Green (yellow) dots mark the positions where the contact resistance in x(y)-junctions peaks and reaches its normal-state value, indicating minigap closure and sign change. Regional minigap closure occurs at the corner in (c) but not in (f).

When the in-plane magnetic field is applied along -45°, the anomalous phase shifts in the two line junctions are the same, so that there is no phase difference jump at the corner. This is evident in Figs. 3(d)-(f) where the behavior of minigap closure at the corner is the same as that in the two line junctions.

To explain the observed phenomena, we need to assume the surface-state EPR of the TIJJ exhibits a 4π-period behavior, as shown by the black lines in Fig. 4(a), causing the minigap to fully close and reopen linearly with a sign change at every odd multiple of π. Applying an in-plane magnetic field at 45° shifts the surface-state EPRs of the two perpendicular junctions in opposite directions, illustrated by the green and yellow lines in Fig. 4(a). These shifts create a parametric region (shaded) where the minigaps in the two line junctions have opposite signs, resulting in a topologically protected regional minigap closure and a zero-energy boundary state at the corner. This state occupies the shaded region in Fig. 4(b), defined by $\text{sign}\left[\cos\left(\frac{\varphi+\varphi_A}{2}\right)\cos\left(\frac{\varphi-\varphi_A}{2}\right)\right] < 0$ [36].

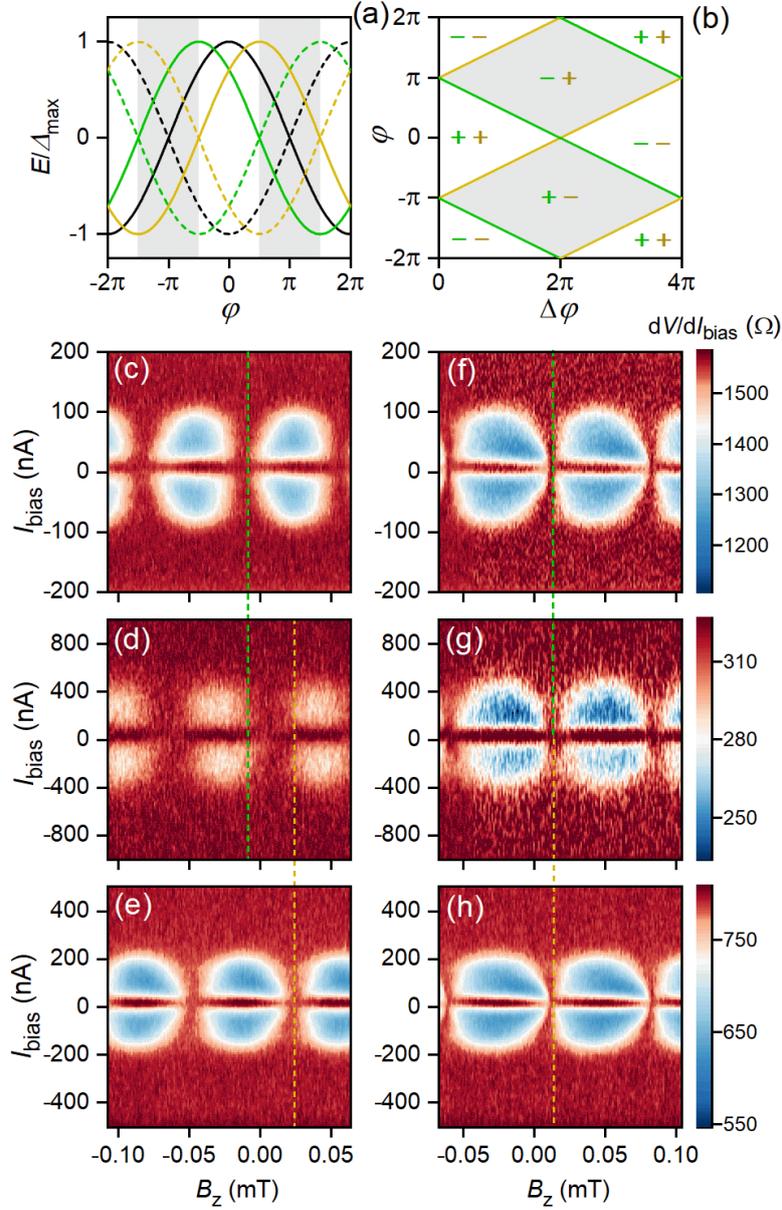

FIG. 4. (a) The 4π-period EPR of the lowest-energy electron-like (dashed lines) and hole-like (solid lines) Andreev bound states in a TIJJ. The minigap is half the energy spacing between them. Applying an in-plane magnetic field at 45° shifts the EPRs of the two line junctions (black lines) in opposite directions (green and yellow lines), creating a shaded parametric region where the minigaps have opposite signs, leading to a topological boundary state at the corner. (b) Phase diagram of minigap closure at the corner under a 45° in-plane magnetic field, plotted against the ordinary phase difference $\varphi$ and the phase difference jump $\Delta\varphi = 2\varphi_A$ at the corner. Green (yellow) lines mark the positions where the minigap in x(y)-junction reaches zero and undergoes a sign change, and the shaded regions denote minigap closure at the corner. (c, d, e) 2D maps of contact resistance at the x-junction center, corner, and y-junction center, respectively, as functions of $I_{bias}$ and $B_z$, with $B_{45°}$=42.4 mT. The minigap at the corner remains closed between the green and the yellow dash lines. (f. g. h) 2D maps of contact resistance at the x-junction center, corner, and y-junction center, respectively, as functions of $I_{bias}$ and $B_z$, with $B_{-45°}$=42.4 mT. No regional minigap closure was observed.

The phase diagram in Fig. 4(b) aligns with the experimental data in Fig. 3(c), both in the regional minigap closure and its 4π periodicity. Although the data in Fig. 3(c) only cover $\Delta\varphi = 0$ to ~2π due to the gradual suppression of the minigap at high in-plane magnetic fields (when the Zeeman energy on the TI surface becomes comparable to the minigap [42]), the minigap amplitude at $\Delta\varphi \approx 2\pi$ (~100 mT), where the superconductivity in the device remains robust [Fig. 3(f)], does not repeat its value at $\Delta\varphi = 0$. This confirms a non-2π-periodic behavior of the minigap, consistent with the 4π periodicity expected in the phase diagram.

In summary, we demonstrated the controllable creation of a boundary state in TI-based Josephson corner junctions, characterized by complete and regional minigap closure at the corner and a deviation from conventional 2π periodicity. These findings are consistent with the Fu-Kane model in which a topologically nontrivial 4π-period Josephson effect is expected, suggesting that TI-based junctions under the Fu-Kane scheme could serve as a promising platform for hosting MZMs. Future work will focus on resolving MZMs by reducing level broadening and suppressing trivial states via methods like electrostatic gating [43], and creating multiple MZMs in more sophisticated device architectures to study the interaction between them.

**References**

1  A. Y. Kitaev, Unpaired Majorana fermions in quantum wires. *Phys.-Usp.* **44**, 131–136 (2001).

2  R. M. Lutchyn, J. D. Sau, S. Das Sarma, Majorana Fermions and a Topological Phase Transition in Semiconductor-Superconductor Heterostructures. *Phys. Rev. Lett.* **105**, 077001 (2010).

3  Oreg, Y., Refael, G. & von Oppen, F. Helical Liquids and Majorana Bound States in Quantum Wires. *Physical Review Letters* **105**, 177002 (2010).

4   Read, N. & Green, D. Paired states of fermions in two dimensions with breaking of parity and time-reversal symmetries and the fractional quantum Hall effect. *Physical Review B* **61**, 10267-10297 (2000).

5   Fu, L. & Kane, C. L. Superconducting Proximity Effect and Majorana Fermions at the Surface of a Topological Insulator. *Physical Review Letters* **100**, 096407 (2008).

6   Lutchyn, R. M. et al. Majorana zero modes in superconductor–semiconductor heterostructures. *Nature Reviews Materials* **3**, 52-68 (2018).

7   Microsoft, Q. et al. InAs-Al hybrid devices passing the topological gap protocol. *Physical Review B* **107**, 245423 (2023).

8   S. Vijay, T. H. Hsieh, L. Fu, Majorana Fermion Surface Code for Universal Quantum Computation. *Phys. Rev. X* **5**, 041038 (2015).

9   Mourik, V. et al. Signatures of Majorana Fermions in Hybrid Superconductor-Semiconductor Nanowire Devices. **336**, 1003-1007 (2012).

10  Deng, M. T. et al. Anomalous Zero-Bias Conductance Peak in a Nb–InSb Nanowire–Nb Hybrid Device. *Nano Letters* **12**, 6414-6419 (2012).

11  Das, A. et al. Zero-bias peaks and splitting in an Al–InAs nanowire topological superconductor as a signature of Majorana fermions. *Nature Physics* **8**, 887-895 (2012).

12  Nadj-Perge, S. et al. Observation of Majorana fermions in ferromagnetic atomic chains on a superconductor. *Science* **346**, 602-607 (2014).

13  Jäck, B., Xie, Y. & Yazdani, A. Detecting and distinguishing Majorana zero modes with the scanning tunnelling microscope. *Nature Reviews Physics* **3**, 541-554, (2021).

14  Dvir, T. et al. Realization of a minimal Kitaev chain in coupled quantum dots. *Nature* **614**, 445-450 (2023).

15  Sasaki, S. et al. Topological Superconductivity in $Cu_xBi_2Se_3$. *Physical Review Letters* **107**, 217001 (2011).

16  Yang, F. et al. Proximity effect at superconducting $Sn-Bi_2Se_3$ interface. *Physical Review B* **85**, 104508 (2012).

17  Xu, J.-P. et al. Experimental Detection of a Majorana Mode in the core of a Magnetic Vortex inside a Topological Insulator-Superconductor $Bi_2Te_3/NbSe_2$ Heterostructure. *Physical Review Letters* **114**, 017001 (2015).

18   Sun, H.-H. et al. Majorana Zero Mode Detected with Spin Selective Andreev Reflection in the Vortex of a Topological Superconductor. *Physical Review Letters* **116**, 257003 (2016).

19   Wang, D. et al. Evidence for Majorana bound states in an iron-based superconductor. *Science* **362**, 333-335 (2018).

20   Liu, Q. et al. Robust and Clean Majorana Zero Mode in the Vortex Core of High-Temperature Superconductor $(Li_{0.84}Fe_{0.16})OHFeSe$. *Physical Review X* **8**, 041056 (2018).

21   Kong, L. et al. Half-integer level shift of vortex bound states in an iron-based superconductor. *Nature Physics* **15**, 1181-1187 (2019).

22   Kwon, H. J., Sengupta, K. & Yakovenko, V. M. Fractional ac Josephson effect in p- and d-wave superconductors. *The European Physical Journal B - Condensed Matter and Complex Systems* **37**, 349-361 (2004).

23   Rokhinson, L. P., Liu, X. & Furdyna, J. K. The fractional a.c. Josephson effect in a semiconductor–superconductor nanowire as a signature of Majorana particles. *Nature Physics* **8**, 795-799 (2012).

24   Wiedenmann, J. et al. 4π-periodic Josephson supercurrent in HgTe-based topological Josephson junctions. *Nature Communications* **7**, 10303 (2016).

25   Lyu, Z. et al. Protected gap closing in Josephson junctions constructed on $Bi_2Te_3$ surface. *Physical Review B* **98**, 155403 (2018).

26   Yang, G. et al. Protected gap closing in Josephson trijunctions constructed on $Bi_2Te_3$. *Physical Review B* **100**, 180501 (2019).

27   Banerjee, A. et al. Local and Nonlocal Transport Spectroscopy in Planar Josephson Junctions. *Physical Review Letters* **130**, 096202, (2023).

28   Microsoft, Q. et al. InAs-Al hybrid devices passing the topological gap protocol. *Physical Review B* **107**, 245423, (2023).

29   von Klitzing, K. et al. 40 years of the quantum Hall effect. *Nature Reviews Physics* **2**, 397-401 (2020).

30   Tanaka, Y., Yokoyama, T. & Nagaosa, N. Manipulation of the Majorana Fermion, Andreev Reflection, and Josephson Current on Topological Insulators. *Physical Review Letters* **103**, 107002 (2009).

31   Nussbaum, J., Schmidt, T. L., Bruder, C. & Tiwari, R. P. Josephson effect in normal and ferromagnetic topological-insulator junctions: Planar, step, and edge geometries. *Physical Review B* **90**, 045413 (2014).

32   Dolcini, F., Houzet, M. & Meyer, J. S. Topological Josephson $\phi_0$ junctions. *Physical Review B* **92**, 035428 (2015).

33   Szombati, D. B. et al. Josephson $\phi_0$-junction in nanowire quantum dots. *Nature Physics* **12**, 568-572 (2016).

34   Assouline, A. et al. Spin-Orbit induced phase-shift in $Bi_2Se_3$ Josephson junctions. *Nature Communications* **10**, 126 (2019).

35   Zhang, X. et al. Anomalous Josephson Effect in Topological Insulator-Based Josephson Trijunction. *Chinese Physics Letters* **39**, 017401 (2022).

36   See Supplemental Material for additional information.

37   Blonder, G. E., Tinkham, M. & Klapwijk, T. M. Transition from metallic to tunneling regimes in superconducting microconstrictions: Excess current, charge imbalance, and supercurrent conversion. *Physical Review B* **25**, 4515-4532 (1982).

38   Plecenîk, A., Grajcar, M., Beňačka, Š., Seidel, P. & Pfuch, A. Finite-quasiparticle-lifetime effects in the differential conductance of $Bi_2Sr_2CaCu_2O_y$/Au junctions. *Physical Review B* **49**, 10016-10019 (1994).

39   le Sueur, H., Joyez, P., Pothier, H., Urbina, C. & Esteve, D. Phase Controlled Superconducting Proximity Effect Probed by Tunneling Spectroscopy. *Physical Review Letters* **100**, 197002, (2008).

40   Beenakker, C. W. J. in *Transport Phenomena in Mesoscopic Systems*, Hidetoshi Fukuyama, Tsuneya Ando. (Springer Berlin Heidelberg, 1991), pp. 235-253.

41   Maistrenko, O., Scharf, B., Manske, D. & Hankiewicz, E. M. Planar Josephson Hall effect in topological Josephson junctions. *Physical Review B* **103**, 054508 (2021).

42   Pientka, F. et al. Topological Superconductivity in a Planar Josephson Junction. *Physical Review X* **7**, 021032 (2017).

43   van Loo, N. et al. Electrostatic control of the proximity effect in the bulk of semiconductor-superconductor hybrids. *Nature Communications* **14**, 3325, (2023).

**Acknowledgements**

This work was supported by the Innovation Program for Quantum Science and Technology through Grant No. 2021ZD0302600; by NSFC through Grant Nos.

92065203, 92365302, 11527806, 12074417, 11874406, 11774405 and E2J1141; by the Strategic Priority Research Program B of the Chinese Academy of Sciences through Grants Nos. XDB33010300, XDB28000000, and XDB07010100; by the National Basic Research Program of China through MOST Grant Nos. 2016YFA0300601, 2017YFA0304700 and 2015CB921402; by Beijing Natural Science Foundation through Grant No. JQ23022; by Beijing Nova Program through Grant No. Z211100002121144; and by Synergetic Extreme Condition User Facility (SECUF).

# Supplementary Material for
# "Controllable creation of topological boundary states in topological-insulator-based Josephson corner junctions"

Xiang Wang[1,2], Duolin Wang[1,2], Yunxiao Zhang[1,2], Xiaozhou Yang[1,2], Yukun Shi[1,2], Bing Li[1,2], Enna Zhuo[1,2], Yuyang Huang[1,2], Anqi Wang[1,2], Zhaozheng Lyu[1,2,3]†, Xiaohui Song[1,3], Peiling Li[1], Bingbing Tong[1], Ziwei Dou[1], Jie Shen[1], Guangtong Liu[1,3,], Fanming Qu[1,3,4] and Li Lu[1,3,4]†

[1] *Beijing National Laboratory for Condensed Matter Physics, Institute of Physics, Chinese Academy of Sciences, Beijing 100190, China*

[2] *School of Physical Sciences, University of Chinese Academy of Sciences, Beijing 100049, China*

[3] *Hefei National Laboratory, Hefei 230088, China*

[4] *Songshan Lake Materials Laboratory, Dongguan, Guangdong 523808, China*

† Corresponding authors: lyuzhzh@iphy.ac.cn, lilu@iphy.ac.cn

**Contents**

1. Device fabrication
2. Creation of a π phase difference jump at the corner by applying in-plane magnetic field along 45° direction
3. The BTK fitting
4. Compensating the out-of-plane component of the "in-plane" magnetic field caused by sample-magnet misalignment
5. Hysteretic response and flux/phase jump caused by large self inductance of the devices
6. The vertical linecuts in Fig. 3(c) of the main manuscript
7. Majorana phase diagram for TI-based Josephson corner junctions

# 1. Device fabrication

Single-crystal $Bi_2Te_3$ was grown using the Bridgman method. Approximately 20 - 30 nm thick flakes were exfoliated from the bulk single crystal using PDMS and transferred onto a $Si/SiO_2$ substrate. The devices were fabricated using conventional electron-beam lithography techniques. The Josephson junctions were formed via superconducting proximity effect by depositing Al/Ti (50/5 nm) superconducting electrodes on the flake using electron-beam evaporation. After the fabrication of the superconducting electrodes, a 30 nm-thick $Al_2O_3$ layer was deposited as an insulating mask. Finally, Au/Al (100/2 nm) was deposited to form the normal-metal probing electrodes for local contact resistance measurement. The tips of the probing electrodes, approximately 100 nm in size, were allowed to contact with the $Bi_2Te_3$ surface. The rest parts of the probing electrodes were deposited on top of the insulating layer. The 2 nm-thick buffer layer of Al in the probing electrodes was not superconducting at the base temperature, and was used to increase the contact resistance between the probing electrodes and $Bi_2Te_3$.

In order to reduce the critical supercurrent of the junction, thus reducing the hysteresis of the rf-SQUID loop (see **Section 5**), the junctions were controlled to be ~1 μm or smaller in size, and the alignment accuracy for different layers was better than 100 nm.

# 2. Creation of a π phase difference jump at the corner by applying in-plane magnetic field along 45° direction

In Fig. 1 of the main manuscript we show that, when the in-plane magnetic field is applied along only one of the two line junctions, the critical supercurrent of the two line junctions are unequally influenced, leading to an incomplete cancellation and thus a non-zero minimum total critical supercurrent of the corner junction at zero perpendicular magnetic field [Fig. 1(d)]. Here we show on another corner junction that when the in-plane magnetic field was applied along 45° direction, the critical supercurrent of the two line junctions was equally suppressed, so that the total critical supercurrent of the corner junction reached zero at zero magnetic field when the phase difference jump at the corner reaches π. The result is shown in Fig. S1.

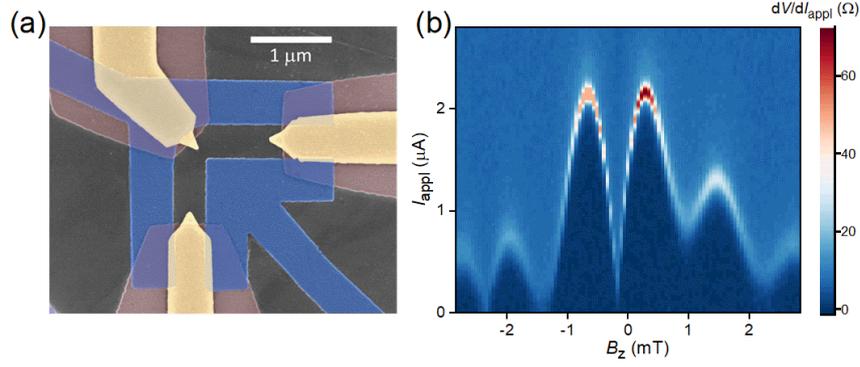

FIG. S1. (a) False-color scanning electron microscope image of another corner junction fabricated on the surface of a $Bi_2Te_3$ single crystal. (b) The Fraunhofer pattern of critical supercurrent measured on another corner junction when the in-plane magnetic field is applied along 45° direction. When $B_{45°}$=53.7 mT, the total critical supercurrent of the corner junction reaches zero at $B_z \approx 0$, implying that the critical supercurrent of the two line junctions are almost equally suppressed by the in-plane magnetic field. Note that the pattern is slightly shifted due to the small earth magnetic field and/or some minor geometric asymmetry of the device.

## 3. The BTK fitting

To map the measured contact resistance to the minigap shown in Fig. 2(e) of the main manuscript, we need a proper single-valued mathematical transformation. Previously, we found that the BTK formalism [37, 38], which was originally developed for analyzing the electron transport processes across the superconductor-normal metal (S-N) interface, could also be effectively applied to treat the electron transport processes across the interface between a normal-metal electrode and the TI surface in a S-TI-S Josephson junction [25, 26]. In this manuscript, we adopted the BTK formalism again to extract the minigap of the Josephson junction.

In Fig. 2(e) of the main manuscript, each data point of the minigap (light gray dots) is obtained through fitting the corresponding vertical linecut in Fig. 2(c) by using the BTK formalism. Here, we provide the details of the fitting.

The BTK formalism is as follows:

$$I_{\text{bias}} = C \int_{-\infty}^{+\infty} [f(E - eV, T) - f(E, T)][1 + A(E) - B(E)] dE,$$

where $f(E)$ is the Fermi distribution function, $A(E)$ and $B(E)$ are the Andreev reflection coefficient and the normal reflection coefficient, respectively, and $C$ is a

constant related to the contact area and the density of states. $A(E)$ and $B(E)$ obey [37, 38]:

$$[1 + A(E) - B(E)]$$
$$= 1 + \frac{\sqrt{(\alpha^2 + \eta^2)(\beta^2 + \eta^2)}}{\gamma^2}$$
$$- Z^2 \frac{[(\alpha - \beta)Z - 2\eta]^2 + [2\eta Z + (\alpha - \beta)]^2}{\gamma^2}$$

where

$$\alpha = \frac{1}{2}\text{Re}\left(1 + \frac{\sqrt{(E + i\Gamma)^2 - \Delta^2}}{E + i\Gamma}\right)$$

$$\beta = \frac{1}{2}\text{Re}\left(1 - \frac{\sqrt{(E + i\Gamma)^2 - \Delta^2}}{E + i\Gamma}\right)$$

$$\eta = \frac{1}{2}\text{Im}\left(1 + \frac{\sqrt{(E + i\Gamma)^2 - \Delta^2}}{E + i\Gamma}\right)$$

$$\gamma^2 = [\alpha + Z^2(\alpha - \beta)]^2 + [\eta(2Z^2 + 1)]^2.$$

And, $\Delta$ represents the phase-dependent minigap in $Bi_2Te_3$, $Z$ is the barrier strength of the $Au$-$Bi_2Te_3$ interface, and $\Gamma$ represents the energy relaxation due to inelastic scattering at the $Au$-$Bi_2Te_3$ interface.

To obtain the detailed values of the minigap shown in Fig. 2(e) of the main manuscript, we applied BTK fitting to the vertical linecuts of the data at each $B_z$ in Fig. 2(c). For example, the fitting to the linecut shown in Fig. 2(d) of the main manuscript yields the dashed black line, with the fitting parameters $\Delta = 48.5$ µeV, $C = 12.4 * \frac{2e^2}{h}$, $Z = 0.7224$, $T = 230$ mK, and $\Gamma = 0$.

We notice that the fitted electron temperature $T$=230 mK is significantly higher than the base temperature of the dilution refrigerator ($T \approx 30$ mK). We believe this is unreal and is mainly caused by the existence of a background (which does not oscillate with varying magnetic field) in the measured contact conductance. Nevertheless, the BTK mapping, though over simplified, still pertinently reflects the linear closing and reopening behavior of the minigap. In fact, the correspondence between the original data of contact resistant and the mapped $|\cos(\varphi/2)|$ shape of minigap in Fig. 2(e) is obvious.

With the obtained fitting parameters, we can also reproduce the measured 2D map of the contact resistance shown in Fig. 2(c) of the main manuscript. The measured and the simulated 2D maps are shown in Fig. S2.

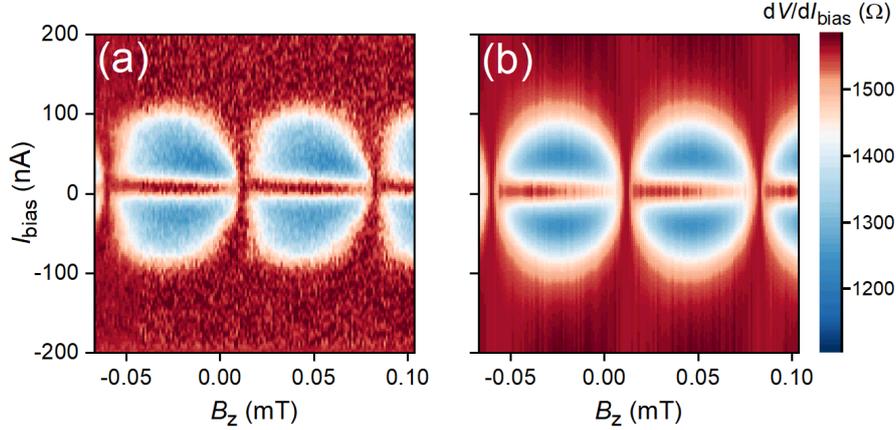

FIG. S2. (a) The measured 2D map of the contact resistance shown in Fig. 2(c) of the main manuscript. (b) The simulated 2D map of contact resistance with fitting parameters $C \in [12,20]$, $Z \in [0.7,1.2]$, $T = 230$ mK, and $\Gamma = 0$.

## 4. Compensating the out-of-plane component of the "in-plane" magnetic field caused by sample-magnet misalignment

Due to the imperfect alignment between the sample and the vector magnets, the applied "in-plane" magnetic field has an out-of-plane component. This component causes unintended flux change in the rf-SQUID loop (thus additional ordinary phase difference across the junction). We performed first-order linear compensation on the out-of-plane components of the x-direction magnet and y-direction magnet individually by simultaneously correcting the z-direction magnetic field $B_z$. In this way, we are able to keep the oscillation of the contact resistance within one oscillation period while varying $B_x$ or $B_y$.

Taking $B_x$ as an example, we varied $B_x$ from zero to 60 mT while measuring the contact resistance at the corner of the device under a bias current of -150 nA, as shown by the blue line in Fig. S3. The contact resistance oscillates over seven periods, indicating that the z-component of $B_x$ at 60 mT is approximately 0.525 mT. Therefore, in subsequent measurements, the applied $B_z$ was corrected by an amount $(0.525/60) \times B_x$. This is a first-order linear compensation on the x-direction magnet. Similar compensation was also performed on the y-direction vector magnet. After such linear compensation, the measured contact resistance no longer oscillates with varying in-plane magnetic field, as shown by the orange line in Fig. S3.

However, such linear compensations are only first-order approximations. It cannot completely compensate the out-of-plane components which often vary nonlinearly due to flux creeping/jumping in the superconducting wires of the magnets. This is why the curves in Fig. S3, as well as the measured patterns in Fig. 3 of the main manuscript, remain irregularly bended.

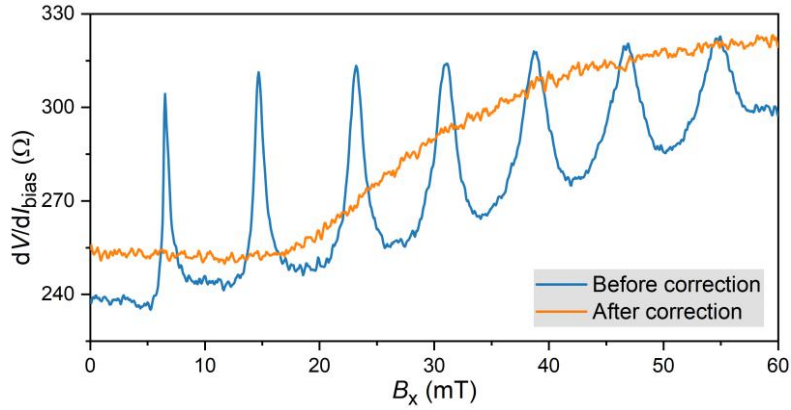

FJG. S3. The measured contact resistance at the corner of the corner junction with varying $B_x$ before and after the linear corrections.

From Fig. S3 it can been seen that the contact resistance does not oscillate to the normal-state value (~320 Ω) at small $B_x$. This is caused by the hysteretic response and flux/phase jumping of the device when self-inductance plays a role at small $B_x$. More detailed explanation can be found in the next section.

## 5. Hysteretic response and flux/phase jump by large self inductance of the devices

When the rf-SQUID loop has a large area and/or a large critical supercurrent, its self inductance plays a role, causing hysteresis and flux/phase jumps in its response to perpendicular magnetic field. Fig. S4(a) shows the 2D color map of the contact resistance as a function of bias current $I_{bias}$ and $B_z$ of device #2. This device has a larger superconducting loop than the one used in the main manuscript, so that it demonstrates more pronounced magnetic flux/phase jumps in the absence of an in-plane field. In Fig. S4(b), the lines of different colors correspond to the vertical linecuts at positions marked by the respective colors in Fig. S4(a). With the exception of the red line which senses the maximum minigap, the lines of other colors sense the minimum values of the minigap at a few $B_z$. It can be seen that: (1) at small $B_z$, the minigap does not undergo complete close due to large hysteresis and sudden jumps; (2) at high $B_z$, the hysteretic and jumping behavior is reduced, so that the minigap undergoes complete close. This is because high $B_z$ modulates and weakens the critical supercurrent of the Josephson junction in the rf-SQUID loop, and reduces the $\beta$ value as explained below.

Taking into account the self inductance of the superconducting loop, the relationship between the magnetic flux inside the rf-SQUID loop $\Phi_{in}$ and the externally applied flux $\Phi_{out}$ obeys: $\frac{\Phi_{in}}{\phi_0} + \frac{\beta}{2\pi}\sin\left(\frac{2\pi\Phi_{in}}{\phi_0}\right) = \frac{\Phi_{out}}{\phi_0}$, where $\beta = \frac{2\pi L I_c}{\phi_0}$ is the hysteresis parameter, $L$ is the self inductance of the superconducting loop, $I_c = I_0 \left|\frac{\sin(\pi\phi_s/\phi_0)}{\pi\phi_s/\phi_0}\right|$ is the critical current of Josephson junction and $\phi_s$ is the magnetic flux threading through the Josephson junction. When $\beta$ is greater than 1, $\Phi_{in}$ becomes a multivalued function with respect to $\Phi_{out}$. With linearly scanning $\Phi_{out}$, $\Phi_{in}$ undergoes phase jump at $\frac{2n+1}{2}\phi_0$ (near the resistance peak) when $\beta$ is larger than 1, as illustrated in Fig. S4(c).

For the device shown in the main manuscript [Fig. 2(a)], its rf-SQUID loop is smaller than device #2, so that the hysteretic and jumping behavior should not be very serious. With increasing in-plane magnetic field, which weakens the critical supercurrent of the junction and reduces the $\beta$ value, the hysteretic and jumping behavior quickly goes away, so that the peak height of the contact resistance oscillations quickly reaches its normal-state value as shown in Fig. S3. The hysteretic and jumping behavior of the devices at low in-plane magnetic fields does not alter the conclusion of the work.

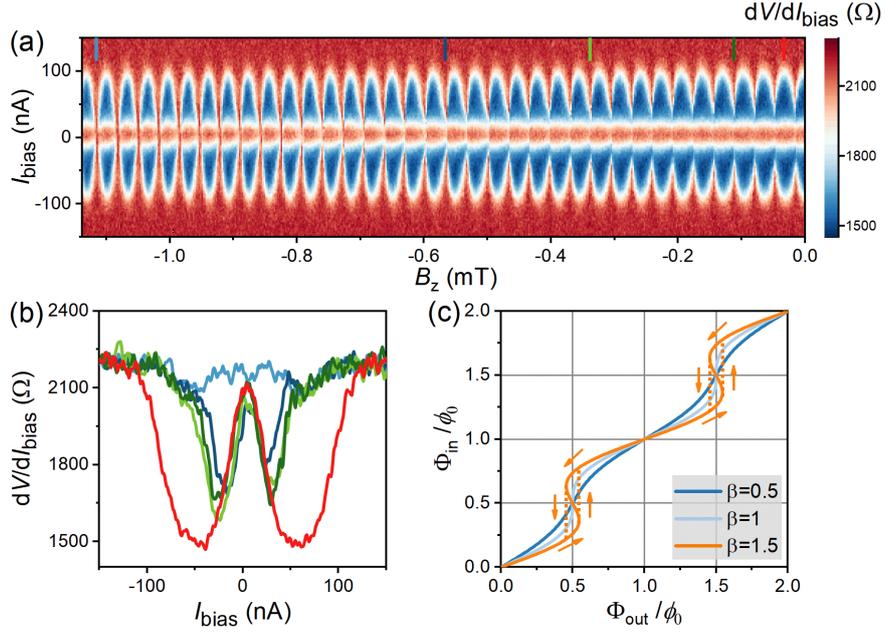

FIG. S4. Hysteresis and phase jump caused by large self inductance in device #2. (a) 2D color map of contact resistance measured at the center of the x-junction as a function of $I_{bias}$ and $B_z$. (b) Vertical linecuts at positions marked with different colors in (a). (c) The applied and effective flux, $\Phi_{in}$ and $\Phi_{out}$, in the rf-SQUID loop with different hysteresis parameter β.

## 6. The vertical linecuts in Fig. 3(c) of the main manuscript

In Fig. S5(a) we replot the measured 2D map of contact resistance shown in Fig. 3(c) of the main manuscript. A few vertical line cuts of the 2D map are shown in Fig. S5(b). The line cuts are taken at $B_{45°}$=30 mT (orange), 60 mT (green), and 90 mT (red). It can be seen from the red line in Fig. S5(b) that the high resistance plateaus really merge together at high $B_{45°}$, demonstrating a complete minigap closure state.

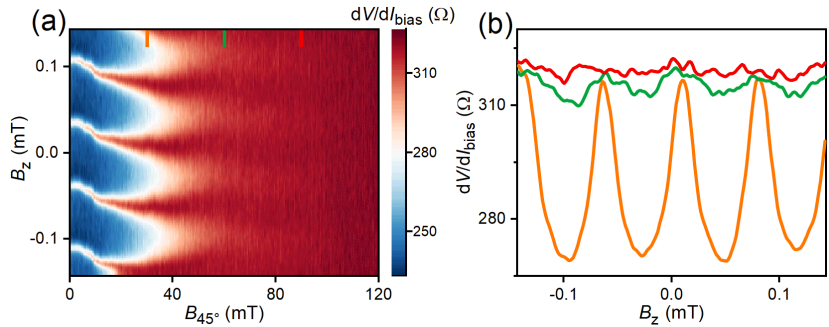

FIG. S5. (a) The measured 2D map of contact resistance at the corner as shown in Fig. 3(c) of the main manuscript. (b) The vertical linecuts at the fields marked in **a** with corresponding colors: 30 mT (orange), 60 mT (green), and 90 mT (red).

# 7. Majorana phase diagram for TI-based Josephson corner junctions

According to Refs. [5] and [41], the Hamiltonian of the TIJJs in the low-energy limit follows the Su-Schrieffer-Heeger (SSH) model: $H = v_e p_\perp \sigma_y + \delta \sigma_z$, where $v_e$ represents the effective Fermi velocity, $p_\perp$ represents the momentum perpendicular to the direction of superconducting phase difference, $\delta \propto \Delta$ is the mass term corresponding to the minigap, and $\sigma_i$ acts on a space composed of two zero-energy modes at $p_\perp = \delta = 0$. $\delta$ is determined by both the ordinary phase difference $\varphi$ and the anomalous phase shift $\varphi_A$: $\delta = \frac{\Delta_{\max}}{\Gamma} \cos\left(\frac{\varphi \pm \varphi_A}{2}\right)$, where $\Gamma$ represents the suppression of minigap by in-plane magnetic field [41]. When the mass term of the two line junctions has opposite sign, the minigap at the corner closes, accompanied by the emergence of a MZM. The Majorana phase diagram of the corner junction when in-plane magnetic field is applied along 45° direction is depicted in Fig. 4(b) of the main manuscript, with the shadowed region defined by $\text{sign}\left[\cos\left(\frac{\varphi+\varphi_A}{2}\right)\cos\left(\frac{\varphi-\varphi_A}{2}\right)\right] < 0$ denoting the minigap closure state.